\documentclass[11pt]{article}

\usepackage[utf8]{inputenc}
\usepackage[round]{natbib}
\usepackage{hyperref}

\usepackage{authblk}

\usepackage{graphicx}
\usepackage{amsmath}
\usepackage{amssymb}
\usepackage{graphicx}
\usepackage{geometry}
\usepackage[colorinlistoftodos]{todonotes}
\usepackage{amsfonts}
\usepackage{amsthm}
\usepackage{indentfirst}
\usepackage{ae}
\usepackage{setspace}
\usepackage{enumerate}
\usepackage{caption}
\usepackage{subcaption}
\usepackage{csquotes}

\usepackage{fancyhdr}
\begin{document}

\title{\bf \LARGE Online Social Network Analysis: A Survey\\ of Research Applications in Computer Science}

\author[1,2]{David Burth Kurka\thanks{d.kurka@ic.ac.uk}}
\author[1,3]{Alan Godoy\thanks{godoy@dca.fee.unicamp.br}}
\author[1]{Fernando J. Von Zuben\thanks{vonzuben@dca.fee.unicamp.br}}
\affil[1]{University of Campinas, Brazil}
\affil[2]{Imperial College London, UK}
\affil[3]{CPqD Foundation, Brazil}

\maketitle

\begin{abstract}
\noindent
The emergence and popularization of online social networks suddenly made available a large amount of data from social organization, interaction and human behavior. All this information opens new perspectives and challenges to the study of social systems, being of interest to many fields. Although most online social networks are recent (less than fifteen years old), a vast amount of scientific papers was already published on this topic, dealing with a broad range of analytical methods and applications. This work describes how computational researches have approached this subject and the methods used to analyze such systems. Founded on a wide though non-exaustive review of the literature, a taxonomy is proposed to classify and describe different categories of research. Each research category is described and the main works, discoveries and perspectives are highlighted.
\\
\\
\textbf{Keywords:} \textit{Online Social Networks, Survey, Computational Research, Machine Learning, Complex Systems}
\end{abstract}

% \begin{keywords}
%  Online Social Networks, Survey, Computational Research, Machine Learning, Complex Systems
% \end{keywords}

\newpage

\section{Introduction}
One of the most revolutionary aspects of the Internet is, beyond the possibility of connecting computers from the entire world, the power to connect people and cultures. More and more the Internet is used for the development of online social networks (OSNs) -- an adaptation of social organizations to the ``virtual world''. Currently, OSNs such as \emph{Twitter}\footnote{\url{https://twitter.com}}, \emph{Google+}\footnote{\url{https://plus.google.com}} and \emph{Facebook}\footnote{\url{https://www.facebook.com}} have hundreds of millions of users \citep{SocialStats}. Futhermore, the average browsing time inside those services is increasing \citep{benevenuto_characterizing_2009} and many websites are featuring some sort of integration with social networking services. Although the effects of such services on personal interactions, cultural and living standards, education and politics are visible, understanding the whole extent of the influence and impact of those services is a challenging task.

The study of social networks is not something new. Since the emergence of the first human societies, social networks have been there forging individual and collective behavior. In the academia, research on social networks can be traced to the first decades of the twentieth century \citep{Rice1927}, while probably the most influential early work on social network analysis was the seminal paper ``Contacts and Influence'' \citep{de1979contacts}, written in the 1950's\footnote{Despite being formally published only in 1978, early versions of this paper circulated among scholars since it was written. These early versions had strong impact on many researchers, including Stanley Milgram in his paper about the small-world phenomenon.}.

In recent years, however, with the popularization of OSNs, this research subject gained new momentum as new possibilities of study have arisen and plenty of data on social relations and interactions have become available. Even though the most popular OSNs have slightly more than ten years of existence -- Facebook was founded in 2004, Twitter in 2006 and Myspace\footnote{\url{https://myspace.com}} in 2003 -- , the volume of scientific work having them as subject is considerable. Finding order and sense among all the work produced is becoming a huge task, specially for new researchers, as the amount of produced material accumulates.

With this in mind, this work aims to present an introductory overview of research in online social network analysis, mapping the main areas of research and their perspectives. A comprehensive approach is taken, prioritizing the diversity of applications, but endeavouring to select relevant work and to analyze their actual contributions. Also, although many disciplines have been interested in this topic -- it is possible to find related works in psychology, sociology, politics, economics, biology, philosophy, to name a few -- , the present work will focus predominantly in computational approaches.

This work is structured as follows: in section \ref{why} the main reasons and motivations for OSN research are discussed; in section \ref{taxonomy} a proposal for a taxonomy is presented and sections \ref{struct}, \ref{data} and \ref{behaviour}, following the proposed nomenclature, detail the main references and findings for each topic. Finally, in section \ref{conclusion} we conclude by presenting general remarks regarding the current stage of the research and a brief analysis of future perspectives.

\section{Online social networks as object of study}
\label{why}

In this section, we make a brief introduction to the research about online social networks, discussing the reasons why this area is getting a very strong momentum, the kind of data being explored in the field and the computational tools commonly used by researchers to analyze social networks data.

\subsection{Why should anyone research OSNs?}

The attention given by the media and general public to OSNs can be a good motivation to justify the research in this field. However, from a computational perspective, OSNs present some particularities that must be taken into account, in order to understand researchers interests. The main reasons are listed below:

\begin{description}
\item[Data availability:] every day, a huge amount of information travels through OSNs and much of it is freely available for researchers\footnote{Respecting, however, specified privacy limits and download rates.}. The current abundance of data has no precedent in the study of social systems and serves as basis for computational analysis and scientific work. Due to its large scale, social data can fit in the context of \emph{big data} research.

\item[Multiple authorship:] differently from other corpora, the textual content produced in OSNs have different authorial sources. This enhances the information content and diversity of the data collected, presenting various styles, forms, contexts and expression strategies. Thereby, OSNs can be a rich repository of text for natural language processing applications.

\item[Agent interaction:] every individual user that composes such networks is an agent able to take decisions and interact with other users. This complex interaction dynamics produces effects that puzzle and interest several researchers.

\item[Temporal dynamics:] the fact that social data is generated continuously along time, allows analysis that take into account spatio-temporal processes and transformations, such as topic evolution or collective mobilization.

\item[Instantaneity:] besides the continuous generation, the social data is also provided at every moment, instantaneously. Thus, OSNs typically react in real time to both internal and external stimuli.

\item[Ubiquity:] following the technological development, which increases people's access to means of communication and information (as smartphones, tablets), OSNs content can be generated, virtually, anywhere and at any time. Also, data's \emph{geolocation}, a feature present in many OSNs, add new possibilities to the analysis.
\end{description}

\subsection{Which networks are explored?}

Two main characteristics can be taken into consideration, before choosing a network to study: popularity (number of active users) and how easy is the data access.

Currently, the largest online social network is \emph{Facebook}, with over one billion active users \citep{fbinfo}. Although the use of data extracted from Facebook is present in literature \citep{Dow2013, Ugander2013, sun_gesundheit!_2009}, the high proportion of protected content -- generally due to users' privacy settings -- severely restricts the analysis using this OSN as source.

\emph{Twitter}, a popular microblogging tool \citep{Cheong2011}, can be considered by far the most studied OSN \citep{rogers_debanalizing_2013}. The existence of a well-defined public interface for software developers\footnote{\url{https://dev.twitter.com}} to extract data from the network, the simplicity of its protocol\footnote{In Twitter, users can post only 140 characters text messages, unlike Facebook, where users can send photos, videos and large text messages.} and the public nature of most of its content can be a good explanation for that. However, since the beginning of the service, rate policies have been created to control the amount of data allowed to be collected by researchers and analysts. This had a direct impact on research, as initial works had access to all the content published in the network, while today's works are usually limited by those policies \citep{rogers_debanalizing_2013}.

It is also worth mentioning the existence of Chinese counterpart services for Facebook and Twitter, like Sina-Weibo\footnote{\url{http://weibo.com}}, the largest one, with more than 500 million registered users \citep{weibo}. Although the usage of those services may differ due to cultural aspects \citep{Yu2011,Gao2012}, similar lines of inquiry can be developed in both the western and eastern equivalents \citep[e.g.:][]{Guo2011, qu_microblogging_2011, Yang2012, Bao2013}.

Other web services that integrate social networking features have been the focus of studies. Examples are media sites like YouTube\footnote{\url{https://www.youtube.com}} \citep{Mislove2007} and Flickr\footnote{\url{https://www.flickr.com}} \citep{cha_measurement-driven_2009, kumar_structure_2010}, and news services as Digg\footnote{\url{http://digg.com}} \citep{hogg_stochastic_2009, wu_novelty_2007}. Research was also made with implicit social networks as email users \citep{tyler_e-mail_2005-1}, university pages \citep{adamic_friends_2003, adamic_how_2005} or blogs \citep{gruhl_information_2004}, even before the creation of social networking services.

\subsection{Computational tools}

There are, currently, many computational tools that help in the task of analyzing large social networks, like graph-based databases (e.g.: AllegroGraph\footnote{\url{http://franz.com/agraph/allegrograph/}} and Neo4J\footnote{\url{http://neo4j.com/}}), libraries to access online social networks APIs (e.g.: Instagram Ruby Gem\footnote{Instagram Ruby Gem is an official Ruby wrapper for Instagram APIs, available at \url{https://github.com/Instagram/python-instagram}.} and Tweepy\footnote{Tweepy is a third-party Python library to access Twitter API. Available at \url{http://www.tweepy.org/}.}), graph drawing softwares (e.g.: Graphviz\footnote{\url{http://www.graphviz.org/}} and Tulip\footnote{\url{http://tulip.labri.fr/}}) and tools for graph manipulation and statistical analysis of networks. The present section, however, will focus only in this last category, as it is more relevant to the kind of analysis conduced in the studies presented in this survey.

Even when considering only tools for graph analysis and manipulation, there are dozens of alternatives, ranging from general purpose graph libraries to advanced commercial tools aimed at specific business. For an extensive list of social networks analysis software, we refer to Wikipedia's entry on the subject\footnote{\url{http://en.wikipedia.org/w/index.php?title=Social\_network\_analysis\_software}, accessed in 16-02-2016}.

When considering applications commonly used in academic works, a division in two groups of tools is clear: (a) graphical user interface (GUI), which are based stand-alone software, focusing on ease of use by non-programmers, and (b) programming language libraries, that are usually more flexible and have more functionalities.

In the first group, the most widely adopted tool is Gephi\footnote{\url{https://gephi.github.io/}} \citep{Bastian2009}, which is a Java-based open source software licensed under the Common Development and Distribution License (CDDL) and GNU General Public License (GPL). Gephi is able to deal with moderate/small graphs (up to 1 million nodes and edges, according to its website), allowing node/edge filtering. It features diverse algorithms to draw graphs, detect communities, generate random graphs and calculate network metrics, like centrality measures (e.g.: betweenness, closeness and PageRank), diameter, and clustering coefficient. It is also able to deal with temporal information and hierarchical graphs and has support for third-party plugins. In addition to the stand-alone software, Gephi is also available as a Java module through Gephi Toolkit\footnote{\url{http://gephi.github.io/toolkit/}}.

Another GUI-based software worth mentioning is Cytoscape\footnote{\url{http://www.cytoscape.org/}} \citep{cytoscape}, also open source and licensed under the GNU Lesser General Public License (LGPL). As Gephi, Cytoscape is written in Java and offers graph drawing, community detection algorithms, network metrics, node/edge filtering and it also supports plugins. Despite being intended for the analysis of biomolecular networks, Cytoscape can be used to analyze graphs from any kind of source, including social networks.

The most adopted and feature-rich libraries in the second group are NetworkX and igraph. Both libraries can handle millions of nodes and edges \citep{akhtar2013analysis} and offer advanced algorithms for networks, as checking isomorphisms, searching for connected components, cliques, communities and $k$-cores, and calculating dominating and independent sets and minimum spanning trees.

NetworkX\footnote{\url{https://networkx.github.io/}} \citep{hagberg-2008-exploring} is an open source project -- under the Berkeley Software Distribution license (BSD) -- sponsored by Los Alamos National Lab, which is in active development since 2002. Despite the recurrent addition of new functionalities, it is a very stable library, as it includes extensive unit-testing. NetworkX is fully implemented in Python and is interoperable with NumPy and SciPy, the language's standard packages for advanced mathematics and scientific computation. It also has remarkable flexibility: nodes can be almost anything -- texts, numbers, images and even other graphs -- and graphs, nodes and edges can have attributes of any type. The library can deal not only with common graphs, but also with digraphs, multigraphs and dynamic graphs. Among the specific features of NetworkX are a particularly large set of graph generators and a number of special functions for bipartite graphs.

igraph\footnote{\url{http://igraph.org/}} \citep{igraph} is a performance-oriented graph library written in C with official interfaces for C, Python and R and a third-party binding for Ruby. If on the one hand it is not as flexible as NetworkX, on the other hand it can be even 10 times faster when performing some functions \citep{akhtar2013analysis}. Many advanced network analysis methods are available in igraph, including classical techniques from sociometry, like dyad and triad census and structural holes scores, and more recent methods, like motif estimation, decomposing a network into graphlets and different algorithms for community detection. As all other tools presented in this section, igraph is an open source project (it is licensed under the GNU GPL).

Two more libraries worth citing are graph-tool\footnote{\url{http://graph-tool.skewed.de/}} and NetworKit\footnote{\url{http://networkit.iti.kit.edu/}}, open source frameworks intended to be much faster than mainstream alternatives by making intensive use of parallelism. Both libraries are implemented mostly in C++ and have Python APIs providing broad lists of functionalities, though not as comprehensive as NetworkX and igraph's. graph-tool \citep{graph-tool_2014}
is licensed under the GNU GPL and is developed since 2006. NetworKit \citep{staudt2014networkit} is more recent: it was created in 2013 in the Karlsruhe Institute of Technology, in Germany. It is under the MIT license and is designed to be interoperable with NetworkX. Differently from other libraries, it aims at networks with billions of nodes and edges and is particularly well-suited for high-performance computing.

The libraries discussed here implement a vast range of graph functions. Some of these functions, however, are not available in all tools. We recommend that researchers in need of specific functionalities to check the libraries' documentation, available at their websites. All these libraries are under active development and are well documented. For more complete comparisons between network libraries, we refer to \citet{combe2010comparative, akhtar2013analysis, staudt2014networkit}.

\section{Categories of study}
\label{taxonomy}

\begin{figure}
\centering
\includegraphics[width=0.7\textwidth]{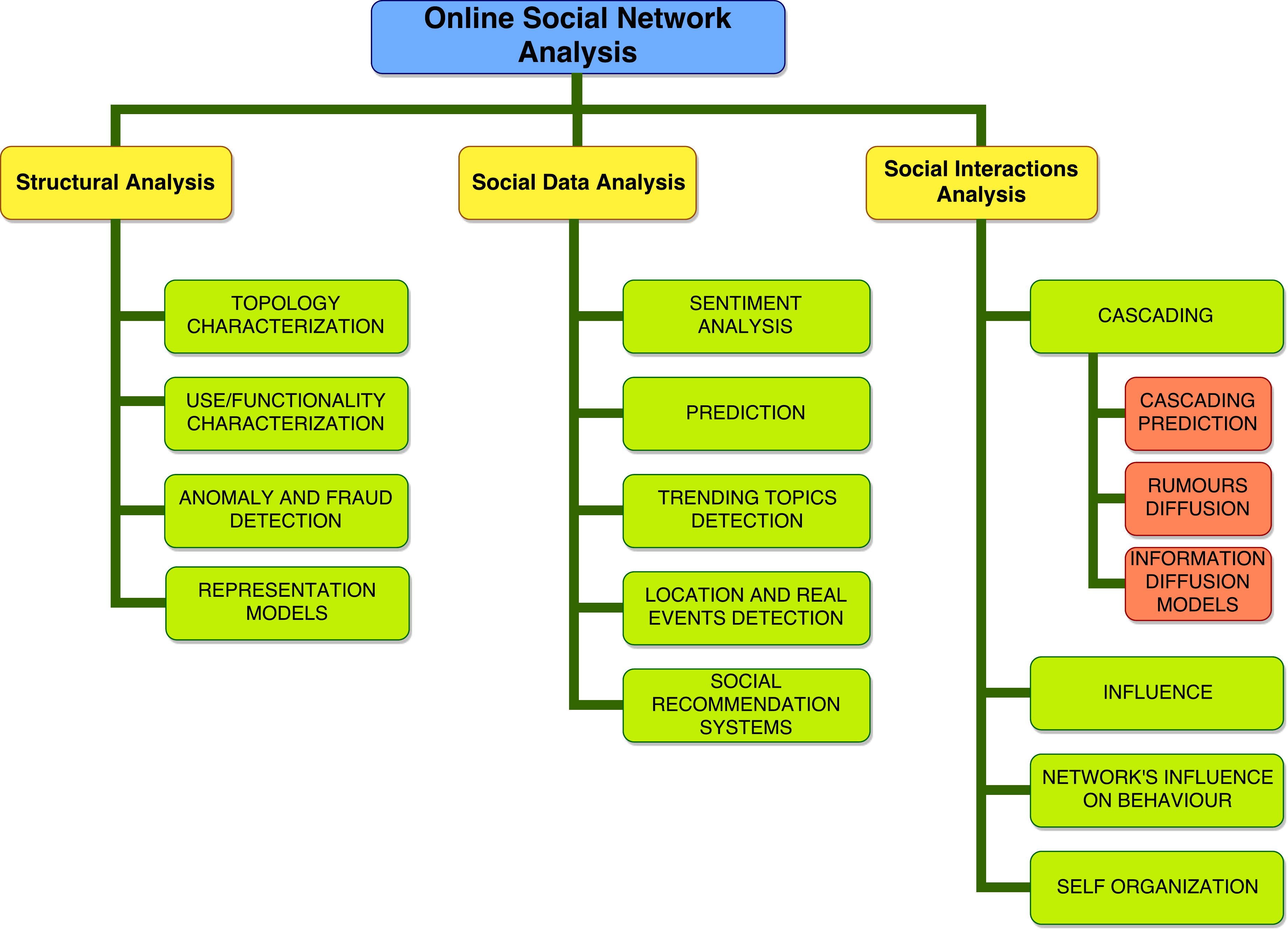}
\caption{Categories of study on Online Social Networks, from a computational perspective.}
\label{fig1}
\end{figure}

In order to simplify the presentation of the wide range of works devoted to the analysis of Online Social Networks, a categorisation of the areas of research is needed. Here we will propose a taxonomy that covers different aspects of this research, structuring all the surveyed works in three main groups: (a) structural analysis, (b) social data analysis and (c) social interaction analysis. Fig. \ref{fig1} illustrates this structure, with its respective subdivisions.

\emph{Structural analysis} is the earliest category of study, since it contemplates initial inquiries about the structure and functionality of social networking services (SNSs), as they were launched. Researchers were interested in simply knowing what are those services and why so many people were being attracted to them. Also, the huge structures that were being formed proved to be worthy investigating and comparing to other known networks (as biological and offline social networks). This area of research is still very active, despite its age.

\emph{Social data analysis} represents a second branch, in which researchers started to use and analyze what OSNs \emph{produce}. This area exploits the huge amount of rich data produced by OSNs to do all kinds of applications. Usually only the data produced by users is considered, not having much importance the topology of users' connections or other network features.

Finally, \emph{social interaction analysis}, deals with aspects related to the individuals using the SNSs. Using all the rich data provided by OSNs, such as users' friendships and the record of social relationships, it is possible to observe how users interact on the network and have insights on aspects of human behavior. This category is intrinsically interdisciplinary, as its discoveries relate to other fields of research, such as psychology, sociology and even biology.

We are unaware of other works that propose a taxonomy for the computational study of OSNs in general. However, previous works were made specifically focusing on studies about Twitter. \citet{Cheong2010} and \citet{Cheong2011} tracked papers produced from 2008 to 2010 and found categories very similar to the ones presented above. However, their general classification is based on only two main areas: user domain and message domain. \citet{Williams2012} systematically collected all the research papers since 2011 containing the word ``Twitter'', and defined four main aspects: message, user, technology and concept. Message could be related to social data analysis, user to social interaction analysis and technology and concept to structural analysis. However, that work did not further deepen the classification in subcategories.

Another interesting perspective is the study conducted by \citet{rogers_debanalizing_2013}, which described the evolution of Twitter and how it has been attracting researchers. According to him, Twitter passed through three phases: Twitter I, when the service was used mainly to connect people, but contained mainly superficial conversations between users; Twitter II, a more mature network, able to promote and organize mobilizations; and Twitter III, a historical valuable big database used to understand society and the recent past.

Of course, we do not expect to achieve consensus with this taxonomy. Imposing categories to any study can be helpful for contextualisation, but can also be misleading and endowed with some degree of arbitrariness. Also, works can belong to more than one category and there can be some intersection between different areas of research. The aim of this survey, therefore, is to serve as an introductory overview of the current status of the field, supported by the proposed taxonomy.

%%%%%%%%%%%%%%%%%%%%%%%%%%%%%%%%%%%%%
%SESSION 1
%%%%%%%%%%%%%%%%%%%%%%%%%%%%%%%%%%%%%
\section{Structural analysis}
\label{struct}

Under structural analysis are works that have OSNs structure and operation as objects of study. Many can be the reasons researchers are interested in the study of a network: to understand how it is composed, to compare its structure to other known networks (specially with offline social networks) or to create models of social organization.

Since the end of the last century, studies showed that many real networks have some non-trivial properties, such as small average distances between nodes\footnote{This is known as the small-world effect, in which the average distance between nodes increases slowly (proportional to $\log N$) in relation to the number $N$ of nodes in the network.} \citep{watts-strogatz-1998} and number of connections per node following a power-law\footnote{In a power-law distribution, the probability of a node to have degree (number of connections) $k$ is given by $p(k) \propto k^{-\gamma}$, where $\gamma$ is a positive constant.} \citep{barabasi-albert-1999}, culminating in the rise of a new area of study named complex networks or network science \citep{Mitchell2009}. Such networks can be found on many areas \citep{Costa2011}, from computer systems to protein interactions and, of course, in social networks. The creation of OSNs and the availability of data, thus, are leveraging this emergent study of complex attributes of OSNs.

\subsection{Topology characterization}

Analyzing the topology of a social network can reveal several interesting features about its components and how people organize themselves for different purposes. Extracting network connections from OSNs is much easier than in offline networks, as all required data is already stored digitally, not asking for explicit knowledge extraction strategies.

Several SNSs had their networks explored and many statistical properties characterized, such as (to name a few):

\begin{itemize}
\item General OSNs services -- Facebook \citep{Ugander2013}, Orkut\footnote{\url{http://www.orkut.com} (defunct since September 2014)} \citep{ahn_analysis_2007, Mislove2007}, Myspace \citep{ahn_analysis_2007}, Cyworld\footnote{\url{http://global.cyworld.com} (defunct since February 2014)} \citep{ahn_analysis_2007, chun_comparison_2008};
\item Media sharing services -- YouTube, Flickr \citep{Mislove2007};
\item Blogging services -- Twitter \citep{huberman_social_2008, Kwak2010}, LiveJournal\footnote{\url{http://www.livejournal.com}} \citep{Mislove2007};
\item Message exchange services -- MSN messenger \citep{leskovec_planetary-scale_2008};
\item Location-based networks -- Foursquare \citep{Scellato2011}.
\end{itemize}

In addition to these services, some studies also attempted to characterize the topology of social networks formed implicitly in sites like university web pages \citep{adamic_friends_2003} and email groups \citep{adamic_how_2005, tyler_e-mail_2005-1}. \\

\noindent
\textit{What the network structure reveals?}

One important property revealed by topology characterization is how similar OSNs are to other real networks previously studied. Agreeing to what is observed in offline social networks, \citet{Mislove2007} verified the presence of power-law degree distribution and small-world property in several OSNs. \citet{Kwak2010} discovered, however, that Twitter's structure does not follow a power-law degree distribution, having an unusual high number of popular users with many followers\footnote{On the Twitter network, connections between users are directional, where one side of a connection is a \emph{follower} and the other a \emph{followee}. Followers receive all the contents posted by the followees, while the reverse is not necessarily true.}, therefore resembling more a news network than a social network.

Using data from MSN Messenger, \citet{leskovec_planetary-scale_2008} analyzed the mean distance between users, identifying small-world property in this network and also showing how people with similar interests (same age, language, location and opposite sex) tend to connect and keep frequent communication. \citet{Ugander2013} discovered that 99.91\% of Facebook users belong to the same large \emph{connected component}\footnote{In a network's connected component, there is a path between each pair of nodes belonging to it. In practice, a huge connected component, like the one found on Facebook, means that almost all users in the network can be reached by any other user in Facebook using only existing social connections.} and that friends communities\footnote{Communities of users can be defined either explicitly, in SNSs where users declare membership to specific groups, or implicitly, as a topological property of the network (which is the case of the article cited here). A topological community is defined by a group of users strongly connected among them, but weakly connected with other groups \citep{Girvan2002a}.} can be stunningly dense, compared to the general sparse structure of the whole network. Also, they showed that common age and nationality are relevant to determine social connections.

The network characterization in services where there is no explicit network allows the inference of interesting discoveries. By characterizing the network formed by internal links connecting web-pages from a university domain, \citet{adamic_friends_2003} showed possibilities of discovering communities and real-world connections among students. From networks built from email services, \citet{tyler_e-mail_2005-1} were able to perceive hidden patterns of collaboration and leadership among users, identifying communities (formal and informal) and leadership roles within the communities. \\

\noindent
\textit{Many networks in one network}

An interesting fact is that an OSN may embed more than one network structure. Many SNSs explicitly register users' relationships, resulting in a \emph{friendship network}. However, from users' interactions, an implicit \emph{interaction network} can also be formed, revealing which social connections are actually active and in use (generally a subgraph of the friendship network). Other possible implicit networks are \emph{diffusion networks}, characterized by the course of a content in the network, and \emph{interest networks}, defined by groups of people with similar interests.

By comparing the friendship network to the interaction network on Twitter, \citet{huberman_social_2008} showed how smaller is the second one, but more adequate to describe and analyze social events. \citet{chun_comparison_2008} showed how Cyworld's interaction network can be more precise to represent real networks, having its nodes' degree distribution closer to known social networks than the friendship network. \citet{wilson_user_2009} discussed that the interaction network can present a different perspective and metrics for an OSN (like larger network diameter and less connected ``supernodes''), being suitable for applications like social spam detection and online fraud detection.

\citet{Smith2014} analyzed conversations on Twitter about different topics and identified, from how participants of a topic are connected, the formation of six distinct network structures according to the subject being discussed. These network structures describe different ``spaces'' of information exchange: from the engaged and intransigent crowds, to the fast content replicating and sharing broadcast networks.

\subsection{Use and functionality characterization}

Since the rise of SNSs, researchers have been interested in understanding the functionality of those services and how their users could take advantage of them.\\

\noindent
\textit{Network formation}

While SNSs were still becoming popular, \citet{Backstrom2006} described how the OSN structure can impact in new friendships and community formation. They showed that more densely connected communities are more likely to receive new members and that events, as the change of the topics of interest in a group, tend to cause transformations in the network topology. \citet{wilkinson_strong_2008} made similar discussion, but focusing on networks of peer production services (Wikipedia\footnote{\url{http://www.wikipedia.org}}, Digg, Bugzilla\footnote{\url{http://www.bugzilla.org}} and Essembly\footnote{\url{http://www.essembly.com} (defunct since May 2010)}), showing how more ancient individuals have a tendency of receiving new connections, concentrating contributions and remaining longer in the network.

\citet{java_why_2007} described, in an introductory perspective, what is Twitter and the main uses of the service: talking about everyday subjects and finding information. Then, they showed how coherent communities arise from the aggregation of users with similar interest. \citet{takhteyev2012geography} analyzed how users' geographical distribution affects their links, uncovering a correlation between the existence of a connection among two users and the frequency of airline flights between the cities they live.\\

\noindent
\textit{User profiles}

Network users can be categorised in different classes by their attributes and patterns of behavior. \citet{krishnamurthy_few_2008} analyzed profiles of almost 100,000 Twitter users and identified three different classes of users: \emph{broadcasters}, with much more followers than followees (e.g.: celebrities); \emph{acquaintances}, with reciprocity in their relationships (e.g.: casual users); \emph{miscreants}, that follow a much larger number of users than they are followed (e.g.: spammers or stalkers).

\citet{Wu2011} identified ``elite'' Twitter users (i.e., celebrities, famous bloggers, media and corporation accounts) and evaluated the impact of the content published by them, realizing that half of the URLs that circulate over the network are generated by 20,000 of those ``elite'' users. Association patterns among those special users are also analyzed, revealing that ``elite'' users of a same field (e.g.: celebrities or blogs) tend to interact among them.

\citet{benevenuto_characterizing_2009} were able to analyze and measure the online activity of users of four SNSs: Orkut, Myspace, hi5\footnote{\url{http://www.hi5.com}} and LinkedIn\footnote{\url{https://www.linkedin.com}}. They discovered that users spend on average 92\% of their time on those services just browsing other users' pages, without posting any content to the network.\\

\noindent
\textit{Conversation}

A notable feature of OSNs is the users' ability to maintain conversations, enabling the organization of mobilization and the creation of enriched content. \citet{Kumar2010} elaborated a detailed study of how conversations are created in diverse OSN contexts, finding patterns and particularities that enabled the creation of a simple mathematical model capable of describing the dynamics of the conversations.

\citet{honey_beyond_2009} analyzed how conversation dynamics can occur on Twitter, with users adapting its simple mechanism of message exchange to track and maintain active communication with each other. In the same line, \citet{boyd_tweet_2010} explored how \emph{retweets}\footnote{A \emph{retweet} is a common practice on Twitter, where a user reposts a message (\emph{tweet}) previously posted by another user, commonly as sign of support or reinforcement.} can be used to create conversations and involve new users in existing conversations.

Discussing the impact of communication in OSNs, \citet{Bernstein2013} discovered, by analyzing large amount of log data, the extent of diffusion of content published on Facebook (i.e., how many people read a message posted by a user). They showed that users usually underestimate the extent of their posts, expecting an audience of less than one third of the actual reached audience.\\

\noindent
\textit{Network deterioration}

Not only the growth, but also the decline in the use of SNSs was studied. \citet{Kwak2011} examined details of the \emph{unfriending} (i.e., unfollowing) behavior on Twitter, showing how frequent it is, using both quantitative and qualitative data, which were obtained through user interviews. \citet{Garcia2013} examined SNSs that suffered intense decline in user activity (Friendster\footnote{\url{http://www.friendster.com} (defunct since June 2015)} and LiveJournal), attempting to understand the impact of users desertion. The impact of ``cascades of users leaving'' on the network resilience was deeply studied, and a metric was proposed to determine when it is or it is not advantageous to users to join a network.

\subsection{Anomaly and fraud detection}

Another important matter that can be explored by structural analysis is the investigation of presence of anomalies and frauds within a network. Those incidents can be either harmless activities, as using false accounts to create artificial number of \emph{likes} in pages \citep{Beutel2013,Jiang2014}, to more serious incidents involving political manipulation \citep{Ratkiewicz2011} or embezzlement \citep{Pandit2007}.\\

\noindent
\textit{Anomaly and fraud}

Analysis of OSN's structure can reveal the presence of anomalies, indicating that users might be acting in suspicious ways.

\citet{Akoglu2010}, observing different networks including email and blogs, examined the topology of sub-graphs formed by users' \emph{1-step} neighborhood. The empirical analysis shows that some properties of those sub-graphs follow a power-law probability distribution, implying that users presenting sub-graphs with improbable values for those properties are considered anomalies and can be inspected. In the presented results, cases such as corrupt CEOs (emails network) or biased connections (blogs network) were detected using the algorithm.

Another interesting example was brought by \citet{Golbeck2015}, who showed that Benford's law -- which predicts the frequency distribution of digits in datasets -- can also be used to detect anomalies in OSNs. It is shown that, in data collected from real SNS, properties such as user's number of posts, number of friends and number of friends-of-friends tend to follow the law. Therefore, the identification of datasets where statistics have a different distribution, can indicate the presence of fraud or suspicious behavior.

A common and practical form of fraud in Facebook's network is the use of automated processes to generate \emph{likes} on the service's pages as a way of artificially promoting a cause, a business or an individual. In order to detect and avoid this situation, \citet{Beutel2013} proposed a method where a bipartite graph is formed connecting users to the pages they liked and registering the time those connections were made. Then, by analyzing patterns of groups of users who liked the same pages, they were able to detect anomalies and misbehavior.

A related problem occurs in some SNSs, where fake accounts are used to increase the number of followers of certain users. \citet{Ghosh2012} investigated this problem in Twitter, analyzing over 40,000 accounts suspended by misconduct. They noticed that, linked to the problematic existence of improper accounts in the service, there are also regular users who, in order to increase their social capital, agree to follow back any user who followed them, creating connections between regular and malicious accounts, hindering the detection of malfunctioning accounts.

Working in the same issue, \citet{Jiang2014} analyzed spatio-temporal properties of OSN sections and created measurements for its `synchronicity' -- how similar and coordinated are the actions of the users on the network section -- and its `rarity' -- how the topology of the section compares to the whole network's structure. The technique was tested in big datasets from Twitter and Sina-Weibo, showing positive results in fraudulent users detection.

Finally, \citet{Jiang2015} summarised many of these techniques and proposed general axioms and metrics to quantify suspicious behavior in OSNs, presenting a new algorithm using these principles which showed improved performance.\\

\noindent
\textit{Spamming behavior}

Another type of fraud occurring in SNS is the presence of user accounts that deliberately send unwanted content (spam) to regular users, abusing the communication channels provided by the services.

This problem in Twitter was tackled by \citet{Benevenuto2010} who identified users acting as spammers in messages related to topics that generated great mobilization. This was made with the use of machine learning techniques that considered the network's structural properties, but also the textual content of messages. \citet{Hu2013a} proposed a related approach, discussing the benefits and challenges of using those features in classification tasks.

Similar problem was also investigated in YouTube. \citet{Benevenuto2009} used properties extracted from the network, users accounts and videos posted, to create a supervised classifier identifying three roles of users: spammers, promoters and legitimates. \citet{OCallaghan2012} worked on the identification of spammers in YouTube's comments using an approach exclusively based on network analysis. For this, a network was built using real data, connecting users to videos, when there was the presence of comments. The formed network's structure presented repeated topology patterns (\emph{motifs}) that, when categorised, lead to the identification of typical structures created by spammer behavior and enabled the systematic identification of suspicious user accounts.

\subsection{Representation models}

One of the challenges of OSN studies is to create models able to describe with success the structure, events and transformations the network goes through. Different models have been proposed addressing this issue. We discuss some of them below.\\

\noindent
\textit{Structure models}

When analyzing the structure of photo sharing OSNs (Flickr and Yahoo!360\footnote{\url{http://360.yahoo.com} (defunct since July 2009)}), \citet{kumar_structure_2010} detected patterns in the network representing different regions: \emph{singletons} (users without connections), \emph{isolated communities} (generally around a popular user) and a \emph{giant component} (users connected to many users). Then, a simple generative model was proposed, able to reproduce the network evolution and recreate the structural patterns observed empirically.

\citet{xiang_modeling_2010} worked on building a model able to represent the \emph{intensity} of social relationships. Instead of having a binary value, each edge between two users in a social graph is calculated as a function of the frequency of interaction among them.\\

\noindent
\textit{Spatio-temporal models}

Although graphs are suitable representations to analyze spatial properties of an OSN, temporal aspects must also be considered in order to represent transformation processes taking place in a network. Although observing temporal aspects of OSN can be a challenge (specially due to the huge number of users involved in processes and data retrieval restrictions), they can be a valuable source of information.

The temporal evolution of a network was studied by  \citet{leskovec_graphs_2005}, who were able to make interesting empirical observations about the growth of several real networks. They noticed that, contrarily to the expectations, the addition of new nodes makes the network become denser in terms of edges per nodes and the average distance between nodes often decreases over time. From those observations, a graph generator model was proposed, able to produce more realistic networks.

\citet{tang_characterising_2010} proposed temporal models to describe network transformations, enabling the creation of new metrics, like temporal distance, i.e., the average time taken for an information published by a user to reach other users. Those metrics are complementary to other spatial metrics (such as geodesic distance) and seem to enable new perspectives of analysis of information diffusion processes or network formation.

%%%%%%%%%%%%%%%%%%%%%%%%%%%%%%%%%%%%%
%SESSION 2
%%%%%%%%%%%%%%%%%%%%%%%%%%%%%%%%%%%%%
\section{Social data analysis}
\label{data}

The focus of social data analysis is essentially the \emph{content that is being produced by users}. The data produced in social networks are rich, diverse and abundant, which makes them a relevant source for data science. As will be seen in this section, most of the computational researches that employ social data use it in machine learning problems such as natural language processing (NLP), classification and prediction. In addition to the challenge of building robust algorithms for such purposes, researchers have also the challenge of building scalable computational solutions that can deal with the large amount of data available in those services.

\subsection{Sentiment analysis}

The textual information produced everyday in SNSs, like Twitter, is a huge corpora \citep{Pak2010}, in which natural language processing techniques, such as sentiment analysis, can be used. Applied to OSNs, sentiment analysis has the potential to describe how emotions spread among populations and their effects.\\

\noindent
\textit{Taking advantage of corpora particularities}

New sentiment classification strategies can be explored, if particularities of the services are taken into account, like the Twitter's (short) size of messages, slangs, \emph{hashtags}\footnote{Hashtag is a text prefixed with the hash (\emph{\#}) symbol. It is commonly used in SNSs to label or tag messages.} and network characteristics. \citet{go_twitter_2009} is one of the first attempts in the literature of sentiment classification on Twitter. Text processing techniques were proposed to extract and reduce features and an algorithm was built reaching over 80\% of accuracy in classification. \citet{Hu2013} also noted that an interesting feature of social data is the presence of \emph{emoticons}, that can be used as labels for machine learning algorithms, helping the process of classification.

Another interesting element of OSN corpora is the presence of language expressions not always present in formal texts. Using the fact that sentences are commonly followed by descriptive \emph{hashtags} (like ``\#irony'' or ``\#not'') that can be used as labels for supervised learning, \citet{Davidov2010} and \citet{Reyes2012} worked on learning and detecting sarcasm and irony in text, with positive results.\\

\noindent
\textit{Applications}

Sentiment analysis can have many applications. For example, \citet{jansen_twitter_2009} and \citet{Ghiassi2013} analyzed how OSN users express sentiments towards different brands, obtaining a measure of approval or disapproval. With the increasing influence of SNSs, this kind of work can be valuable for companies to understand and deal with customer demands. \citet{dodds_temporal_2011} and \citet{lansdall-welfare_effects_2012} developed indicators of happiness among populations, based on the analysis of OSNs texts. With that, they were able to analyze the impact of historical events -- such as economic recession \citep{lansdall-welfare_effects_2012} -- in public opinion, showing an innovative quantification of population welfare.

Deeper analyzes take into account not only text classification, but also a study of how sentiment spread in the network. \citet{Hu2013Exploiting} took advantage of \emph{emotional contagion} theories \citep{Rapson1993} to help the classification of texts produced by specific users, having better results than traditional algorithms. In a controversial experiment, \citet{Kramer2014} filtered content displayed on Facebook to emphasise positive or negative posts, showing how emotions can be contagious. Although the users subject to the experiment did not presented drastic changes of behavior, there were statistically significant effects observed.

\subsection{Prediction}

A valid question to address when dealing with OSNs is how representative are the dynamics present in the virtual environment in relation to the non-virtual world. Supposing that what happens inside SNSs can provide information about other external events, researchers have been trying to build predictors in many fields:

\begin{itemize}
\item Elections: predict the outcome of elections from OSNs manifestations \citep{tumasjan_predicting_2010};
\item Box-office revenue: forecast the popularity (and revenue) of a blockbuster before or just after it comes out \citep{asur_predicting_2010};
\item Book sales prediction \citep{Gruhl2005};
\item Disease spread \citep{culotta_towards_2010, lampos_nowcasting_2012};
\item Stock market prediction from sentiment analysis \citep{bollen_twitter_2011}.
\end{itemize}

However, despite the initial positive results and good perspective presented in the works above, skepticism about the effectiveness of the proposed methods and their representativeness must be noted, as seen in \citet{gayo-avello_meta-analysis_2013}, \citet{Wong2012} and \citet{zhang_predicting_2011}, which analyzed election forecasts, box-office revenue and stock market predictions, respectively. Those studies showed that the validity of the initial findings can be questioned and that many results can not be generalized as expected.

\subsection{Trending topics detection}

Another important focus of research that uses content published in OSNs is the analysis of message exchange dynamics, aiming to detect trends. Although some SNSs, like Twitter, have their own algorithms for trending topics detection, alternative proposals of content detection and organization have been made.
According to \citet{Guille2013}, there are two main approaches to detect a trending topic in an SNS: message analysis or network analysis.\\

\noindent
\textit{Message analysis}

Focusing on the messages content, \citet{Shamma2011} proposed a simple metric to identify trending topics, analyzing the frequency of words during specific time frames, compared to its general frequency (similar to the usual tf-idf \citep{Salton1986} model in NLP). A trending topic happens when there is an abnormal term frequency occurrence.
In a creative approach, \citet{weng_event_2011} considered the frequency in time of words as waveforms. Thus, some messages would contain words with waveforms that resonate together, enabling the identification of emergent topics.

\citet{Lu2012} went beyond and developed a method to predict which topics will be popular in the future. Using strategy originally intended to predict stock markets, this method is able to calculate the \emph{trend momentum}: the difference of frequency of a term between a short and a long time period. In the tests performed, the method was effective, with trends being successfully predicted by the increase of the momentum.\\

\noindent
\textit{Network analysis}

On the transition from message to a network approach, \citet{cataldi_emerging_2010} used not only the term frequency, but also the authority (calculated using PageRank) of users posting the observed content. This way, they were able not only to identify trending topics, but also related topics. \citet{takahashi_discovering_2011} used exclusively network information to create a probabilistic model of interactions. When anomalies are detected in the interaction pattern, a trending topic can be detected, without even text analysis. In their tests, this technique performed at least as good as other text-based techniques, being superior when topic keywords are hard to determine.\\

\noindent
\textit{Tracking memes evolution}

Apart from trending topics detection, \citet{Leskovec2009} studied not only topics created, but also their evolution in new subtopics or derivatives over time, observing the spreading of news for days. The researchers were able to track a common path in the news cycle, with content being first published in traditional media and, few hours later, the same content appearing in blogs and other online services, resulting in ``heartbeat-like'' patterns of attention peaks.

% In the same line of meme detection, \citet{Kuhn2014} \todo{Esse artigo, aparentemente, não tem nada a ver com redes sociais online. Precisa ou tirar ou mostrar como ele se liga com OSNs.} analyzed the evolution of scientific ideas, identifying repeated terms and the connections in the network of citations between scientific papers. Thus, they developed a ``meme score'' from terms' frequency and propagation degree, enabling the identification of relevant concepts within a scientific field.

\subsection{Location and real events detection}

In many cases, topics discussed in OSNs are about events that take place in the ``real'' (or external) world, like political, public or daily life events. Also, as contents are often posted from mobile devices, it is common for OSN users to be physically present during those events. Therefore, OSN data can be a valuable resource for recovering data from offline interactions.\\

\noindent
\textit{Location}

Information about geographical localization of OSN users is available in many SNSs, specially in \emph{location-based SNS}, such as Foursquare\footnote{\url{https://www.foursquare.com}} and Nearby\footnote{\url{https://www.wnmlive.com/}}.
\citet{Noulas2011} characterized users' geographical data present on Foursquare, demonstrating the potential of such data in unprecedented research on human mobility, urban spatiality and in applications such as recommendation systems.

\citet{Cho2011}, analyzing social data from both location-based SNS (Gowalla\footnote{\url{http://www.gowalla.com} (defunct since March 2012)} and Brightkite\footnote{\url{http://www.brightkite.com} (defunct since April 2012)}) and from cell phone towers, found patterns on user mobility, being able to create a predictive model of users location. The analysis reveals that, although people tend to stay most of the time transitioning between routine locations (e.g.: home, work), social connections and location of friends are also determinant to identify an individual's location.

The geolocation of an OSN user can also influence its relationships and content exchanged. \citet{takhteyev2012geography} found that groups of people sharing similar cultural or geographical elements, such as language and location, are more likely to be connected in an OSN. Also, the existence of physical connections between places, like the presence of abundant airline flight routes, can be an indicator of social connections. \citet{Cheng2010} also explored this aspect, indicating that it is possible to predict the location of a user exclusively from the content of his/her textual messages, even when this information is not explicitly disclosed.

Showing the potential of social data as a demographic tool, \citet{Cranshaw2012} developed a methodology where a city can be spatially divided in regions, using data from Foursquare. By comparing the record of users present in different public spaces (Foursquare's \emph{check-ins}) and the spaces' geographical locations, an affinity matrix is built, revealing similarities between premises. This matrix can then be clustered, revealing areas of both spatial and social proximity inside cities. These areas, denominated by the authors as \emph{livehoods}, form a relevant and coherent territory demarcation (as revealed by interviews), presenting as a valuable alternative to traditional municipal organizational units such as neighborhoods.\\

\noindent
\textit{Detecting real events}

\citet{Becker2011} worked on a method to distinguish Twitter messages that refer to real events from those that do not (jokes, spam, memes, etc.) by clustering messages of the same topic and, then, classifying the clusters based on their properties. \citet{Psallidas2013} discussed the challenge of separating, in an OSN, content related to predictable events (e.g.: awards, games, concerts) from those related to unpredictable ones (e.g.: emergencies, disasters, breaking news). Features useful to describe each type of diffusion were evaluated to be used as input to classification algorithms, being effective in large-scale experiments.

\citet{Sasahara2013} analyzed how some topics related to past events spread across the social network, finding some patterns that help in the identification of real event diffusion. According to the authors, diffusion networks of real events have an abrupt and unusual structure (compared to diffusion of other kinds of events), making it possible to create automatic tools to detect them.\\

\noindent
\textit{Using real events information}

\citet{hu2012breaking} studied how a social network is capable of disclosing breaking news even before traditional media. They used as case study the fact that the news of Osama Bin Laden's death were disclosed on OSNs before traditional media and showed how OSN users take roles of leadership to efficiently transmit information and influence other users on those events.

Using this ability of quick awareness, \citet{Sakaki2010} and \citet{neubig_safety_2011} proposed automatic methods for detecting earthquakes in Japan, considering network users as \emph{social sensors}. Their results were robust and promising, involving the identification of the earthquake's centre and trajectory, inference about the safety of people possibly affected and the generation of automatic earthquake alerts faster than official announcement by authorities.

\subsection{Social recommendation systems}

Another application of OSN data is the possibility of creating social recommendation systems for products or even content produced by users in the network. In a space with many users and data, the use of social relationships can improve traditional recommendation systems both in relevance and scalability, as users connected by social relationship usually share many interests, both by homophily\footnote{The tendency of an OSN user to connect to similar people.} and by contagion, reducing the amount of data necessary to make accurate recommendations.\\

\noindent
\textit{Trust networks}

One practical use of social information in recommendation systems is the synthesis of \emph{trust networks}, which are groups of related users that are considered to have a valuable opinion on some matters. Generally, a user's truthfulness is related to its proximity to a reference user.

\citet{walter_model_2008} described how an OSN can be used to collect information in general and how the relationships can help to filter relevant information for each user, as trust networks are established. By using exclusively content on users' neighborhoods, they were able to build effective recommendation systems as good as other systems that use information from the whole database. \citet{Arazy2009} created social recommendation systems in order to evaluate products reputation, building trust networks to ponder the relevance of users opinions.\\

\noindent
\textit{Improving traditional recommendation systems}

Other uses of OSN data for recommendation systems include the work of \citet{Ma2011}, who uses relationship data to initialize recommendation systems that have few initial reviews. Also, \citet{Yang2013} created probabilistic models to model users preferences and make recommendations based on friendship connections. In a more conservative proposal, \citet{Liu2010} suggested ways to improve existing recommendation systems by including social information, like users' relationships, and showed how the accuracy of algorithms may be positively affected.\\

\noindent
\textit{Content selection}

A common task for recommendation systems on SNSs is to select relevant content to be displayed to users. \citet{chen_short_2010} worked on a series of algorithms to recommend content to users, in order to improve Twitter's usability. They were able to reach a level in which 72\% of the content showed was considered interesting, according to real Twitter users feedback.

\citet{Backstrom2013} worked with Facebook data, analyzing the attention a topic might receive, by predicting the topic's length and its re-entry rate (i.e., the number of times a user participates in the same topic). This gives a measure of how interesting a topic is and can be used to select and recommend content to users.

%%%%%%%%%%%%%%%%%%%%%%%%%%%%%%%%%%%%%
%SESSION 3
%%%%%%%%%%%%%%%%%%%%%%%%%%%%%%%%%%%%%
\section{Social interaction analysis}
\label{behaviour}

By watching \emph{users diffusing content}, there is the expectation of knowing more about complex human behavior. The access to data produced by OSNs and the knowledge of how to process and analyze them are enabling computer scientists to join discussions previously exclusive to sociologists or psychologists. This new intersection of fields is known as \emph{computational social science} \citep{Lazer2009, CioffiRevilla2010, Conte2012}.

There are still questioning related to whether the behavior observed in an OSN can be extrapolated to its users offline lives and whether OSN users are representative enough for drawing conclusions, from their behavior, for whole societies \citep{Boyd2010}. Even so, there is a plenty of phenomena that take place on OSNs that are worth to be studied, as we will outline in this section.

\subsection{Cascading}
\label{cascading}

One of the most widely studied behavioral phenomenon that takes place in OSNs is information cascade. Also known as viral effect, a cascade is characterized by a contagious process in which users, after having contact with a content or a behavior, reproduce it and influence new users to do the same. This decentralized process often causes chain reactions with great proportions, involving many users and being one of the main strategies for information diffusion in social networks.

The unpredictability and the magnitude of this phenomenon attract many researchers, trying to interpret and understand the factors behind it. The cascade effect has been studied and characterized in many different SNSs, as:

\begin{itemize}
\item Facebook \citep{sun_gesundheit!_2009, Dow2013};
\item Google+ \citep{Guerini2013};
\item Second Life\footnote{\url{http://secondlife.com}} \citep{Bakshy2009};
\item Flickr \citep{cha_measurement-driven_2009};
\item Twitter and Digg \citep{Lerman2010};
\item LinkedIn \citep{Anderson2015}.
\end{itemize}

\citet{Goel2012} alone studied information diffusion in seven different OSN domains, verifying similarity in cascading properties, regardless the service observed.\\

\noindent
\textit{Properties observed}

From the empirical analysis of information cascades on OSNs, some common properties can be observed, as already shown by \citet{Goel2012}. A good characterization of many of those properties can be found in \citet{borge-holthoefer_cascading_2013}, that gathered results from works that modeled and analyzed cascades.

Among the properties observed, some are highlighted:

\begin{itemize}
\item Most cascades have small depth\footnote{The depth of a diffusion network (or tree) is the maximum distance between the diffusion source (the root) and the users involved in the diffusion. A distance between two users is defined as the size of the shortest path on the network that connect them.}, exhibiting a star-shaped connection graph (a central node connected to many others around it). This was shown by many researchers, as \citet{leskovec_cascading_2007}, \citet{Gonzalez-Bailon2011}, \citet{Lerman2010} and \citet{Goel2012}.
\item In practice, the majority of information diffusion processes that take place in the network are shallow and do not reach many users. Thus, widely scattered cascades turn to be rare and exceptional events.
\item In general, cascades (even large ones) occur in a short period of time. Most reactions to a content posted on an OSN usually happen quickly after it is posted \citep{Centola2010, leskovec_cascading_2007} and do not last for a long time \citep{borge-holthoefer_cascading_2013}.
\item Any user on the network has potential to start widely scattered cascades. It is shown that different sources of information can conquer space on the network  \citep{Mocanu2014}, and attempts to measure users' potential to start a cascade are not conclusive \citep{Bakshy2011, Borge-Holthoefer2012} (see section \ref{influence} on influence for more details).
\end{itemize}

\noindent
\textit{Information origins}

\citet{Myers2012} studied sources of information in OSNs. They found that almost one third of the information that travels on Twitter network comes directly from external sources, while the rest comes from other users, through cascades. Tracking a cascading process can be a challenge when the content being propagated may undergo changes. \citet{Leskovec2009} proposed ways to track memes and their derivatives, in a process that can take several days, showing the long transformation process from publication to popularization.\\

\noindent
\textit{How topology influences cascades}

The analysis of the network underlying a diffusion is a helpful way to understand a cascading process. \citet{Goel2013}, using a dataset of billions of diffusion events on Twitter, analyzed the diffusion networks and proposed a ``structural virality'' metric, able to measure the network's tendency to successfully propagate an information.

One of the most important conclusions of the network analysis, shown by \citet{sun_gesundheit!_2009}, \citet{Ardon2011} and \citet{weng_virality_2013}, is the fact that topics that can reach initially more than one community of users tend to cause larger cascades.\\

\noindent
\textit{Cascades from historical events}

Specific events where SNSs had significant influence, such as political movements and protests, received special attention in social network analysis. In 2009, following the Iran presidential elections, many protests took place and their effects could be noticed in SNSs by increased diffused information. \citet{zhou2010} conducted a qualitative research of these cascades, concluding that in general they are shallow (99\% of the diffusion trees have depth smaller than three). \citet{Gonzalez-Bailon2011}, based on the diffusion network, analyzed the roles of users and related them to their positions in the network. According to the study, influential users in the process of spreading information tend to be more central in the network.

Similar experiments were made with protests that happened in Spain on May 15th 2011. \citet{borge-holthoefer_structural_2011} analyzed the diffusion network related to such events and differentiated users that acted as sources of information and users that only consumed it. In a later work, \citet{gonzalez-bailon_broadcasters_2013} identified four types of users -- namely influentials, hidden influentials, broadcasters and common users -- that can help the understanding of how users behave in cascading processes.

\subsection{Predicting cascades}

An important motivation for characterizing cascades is to be able to predict how users in a network will behave with regards to a specific content and how this content will spread. This capacity to tell beforehand how many users will see or share an online content can be a source of revenue for advertisers and, also, a useful tool to governments willing to effectively disseminate public interest information.

However, the task of predicting popularity of online content has shown to be extremely difficult to accomplish \citep{salganik_experimental_2006, Watts2012}. Two main problems are determinant \citep{Cheng2014}: (a) the definition of what are the features (if any) that determine the size of a cascading process; and (b) the fact that widely spread cascades are rare events \citep{Goel2012}, making it hard to develop and train algorithms with so few positive samples.

Nevertheless, those difficulties were not enough to prevent research in this area, as seen in the many scientific works already published. Also, according to experiments presented by \citet{Petrovic2011}, the identification of content likely to be shared is a task manageable by humans, what can bring hope to new inquiries. As we will show below, many are the works published in this topic and so are the strategies used to tackle the problems.\\

\noindent
\textit{Feature selection}

The most important aspects to be considered when building machine learning algorithms (such as predictors or classifiers) to analyze cascades is the proper characterization of information diffusion processes and the choice of relevant properties to describe these processes preserving existing distinctions among them \citep{Suh}. From the literature, we can see that four main classes of features are generally chosen: (a) message features, (b) user features, (c) network features, and (d) temporal features.\\

\noindent
\textit{Message features}

Does a textual message posted in an OSN have an intrinsic potential to be shared? Assuming that some content has more potential than others to create cascades, researchers have investigated ways of predicting the future popularity of a message based on text analysis. This kind of investigation might be specially interesting in cases in which there is the need (or the will) of maximizing the audience reached by a content posted by a specific user. Thus, by adjusting the text that will be posted, it would be possible to increase the range of an author's message.

This is the aim of \citet{Kunegis2011} work, that found correlations between message content and \emph{retweet} count on Twitter. Several features were analyzed, such as presence of URLs, \emph{hashtags}, mention to other users, punctuation and sentiment analysis. Their conclusion is that messages referring to public content and with negative emotions are more likely to be shared. \citet{Suh} did an extensive search for features, both in message and user characteristics, in a large dataset (74 million posts from Twitter) highlighting the presence of URLs and \emph{hashtags} as the most relevant factors in the message content for predicting cascades.

More creative message descriptors were studied by \citet{Hong2011}, who used topic detection algorithms to identify a message's topic, to be further used as a feature. \citet{Tsur2012} explored different interesting features that can be extracted from a \emph{hashtag}, like its location inside a post or its size in characters or words.\\

\noindent
\textit{User features}

It is evident that a popular and influential user has more chance of generating a cascading process than an anonymous user. Therefore, analyzing aspects related to the user that shares a message, and possibly about the users that continue this process, can be crucial to build a reliable cascade predictor.

In addition to message features (as discussed above), \citet{Suh} also analyzed a set of possible features related to authors, including the number of connections, number of past messages posted, number of days since the user's account was created and number of messages previously marked as favorite by other users. Their conclusion was that only the number of connections and the age of the account have any sort of correlation to \emph{retweet} rates. \citet{Hong2011} also suggested other features, namely: author's authority according to PageRank \citep{page_pagerank_1999}, degree distribution, local clustering coefficient\footnote{Clustering coefficient is a measurement of network cohesiveness. The local clustering coefficient for a specific node is given by the number of direct connections between two of its neighbors, divided by the number of possible connections between these neighbors.} and reciprocal links.

Metrics taking into account properties of the users involved in a diffusion (beyond the author) can be also valuable. \citet{Hoang2012} introduced a model to predict information virality on Twitter, by creating three features: item virality (the rate of users that share a content, after receiving it), user virality (the number of connections of users involved in a diffusion) and user susceptibility (the proportion of content shared in the past by a user). \citet{hogg_stochastic_2009}, by observing cascades on Digg, were able to create models that describe the initial behavior of users sharing content, thus allowing the forecast of a cascade's size. \citet{Lee2014} explored features related to previous behaviors of users, such as average time spent online, time of the day in which the user is more likely to join discussions, and number of messages sent over time.\\

\noindent
\textit{Network features}

The analysis of the network structure where a diffusion takes place is also important to determine the potential range of a cascade.

\citet{weng_virality_2013} explored the importance of a network characterization, using the knowledge that diffusions starting in multiple communities are more likely to be larger \citep{sun_gesundheit!_2009, Ardon2011}. The authors then proposed as a metric the number of communities involved in the early diffusion and the amount of message exchanges between different communities (inter-community communication).

\citet{Kupavskii2012} examined a set of features to describe a cascade, showing relevant improvements in the prediction task when using network features such as the flow of the cascade -- a measure related to the number of users sharing a content and how fast they share it -- and the authority in the network formed by users sharing the same message, calculated using PageRank \citep{page_pagerank_1999}. \citet{Ma2013} used both message and network features to predict the popularity of Twitter \emph{hashtags}. Among the network features adopted are metrics like the ratio between the number of connected components in a network and the number of users that initiated the cascade, the density of the diffusion network\footnote{The density of a network is the ratio between the number of actual connections and the number of possible connections.} and the diffusion network's clustering coefficient. Their conclusion is that network features are more effective than message features for predicting the use of \emph{hashtags}.\\

\noindent
\textit{Temporal features}

Every cascade process can be represented as a time series, listing the amount of information diffused over time. This time series can be seen as a cascade signature, representing its range, speed and power.

\citet{Szabo2010} analyzed the initial diffusion of YouTube and Digg contents and, based on the initial time series, forecast the long term popularity of specific contents. They pointed that only two hours of data about the access to Digg stories was enough to predict thirty days of popularity, while, on YouTube, ten days of records were needed to evaluate the next twenty days.

\citet{Cheng2014} improved this strategy, by dividing the original prediction problem into subtasks where, based on past features, a classifier must estimate if a content published on Facebook will double its audience or not. Thus, robust and high performance classifiers can be built.\\

\noindent
\textit{What exactly is predicted}

After presenting the features used to describe cascading phenomena, it is worth examining the different approaches to predict cascades.

Most of the work in this topic tries to measure the number of users or messages that will join a cascade. Examples are \citet{Kupavskii2012}, who worked predicting the number of messages (\emph{retweets}) a cascade will have, \citet{Ma2013}, that predicted the popularity of a new topic (\emph{hashtag}), and \citet{Suh}, that forecast the rate of users participating in a cascade.

However, some works were simply interested in building binary classifiers to determine if a content will be shared by any user or not. This is the case of \citet{Kunegis2011} and \citet{Petrovic2011}.  \citet{Hong2011} went a little further and created four categories of cascading -- not shared, less than 100 shares, less than 10000 shares and above 10000 shares -- that can be classified more easily.

Another strategy was used by \citet{Galuba2010}, who built a system able to predict which users are leaned to enter a cascade. \citet{Lee2014} worked in the same line, being able to sort the $N$ users most inclined to share a message.

\subsection{Rumors diffusion}

Another particular area of study involving cascading that received special attention from the research community is the detection of false information (rumor\footnote{Although the word ``rumor'' is used in this work exclusively with the sense of false information, some areas of the literature might also use it to refer to information in general \citep[e.g.:][]{daley1965}}) propagation.\\

\noindent
\textit{Characterizing rumors}

Aiming to characterize this phenomena, \citet{Friggeri2014}, with the assistance of a website that documents memes and urban legends (\url{http://snopes.com}), mapped the appearance of rumors on Facebook network, showing that rumor cascades tend to be more popular than generally expected and discussing users' reactions after acknowledging the falsehood of previously posted messages. Also on Facebook, \citet{Mocanu2014} observed the acceptance by network users of different sources of information. By analyzing how content from (a) mainstream media, (b) alternative media, and (c) political activism is diffused, they concluded that, regardless of source, every information has the same visibility. This may favour people that share false content, as they potentially have the same power of influence on the network as reliable sources.\\

\noindent
\textit{Detecting rumors}

 \citet{Mendoza2010}, when analyzing the diffusion networks of news related to a natural disaster in Chile, realized that the patterns of rumor spreading are different from those related to real information spreading. Therefore, in a subsequent work, \citet{castillo2011information} sought automated methods to detect rumors, by analyzing features from texts posted and the users involved in the propagation of the information.

\citet{qazvinian2011rumor} further proved the effectiveness of using features related to network and message content to detect rumors. Despite their positive result, it is noticeable the small number of rumors analyzed (only five), given the quantity of data (10000 posts from Twitter). \citet{Gupta2012} also worked developing metrics, but this time trying to measure credibility of users, messages and events, resulting in a score for the credibility of the general topic diffused.\\

\noindent
\textit{Rumor containment}

In a different perspective, \citet{Tripathy2010} explored ways to contain a rumor cascade, after its identification. Using techniques inspired by disease immunization, they discussed the importance of a quick identification of rumors and the use of anti-rumors agents able to detect such events and spread messages against the rumors. Lastly, \citet{Shah2010} aimed to detect the source of a rumor cascade, developing a new topological measure entitled ``rumor centrality'', able to outperform traditional metrics in special cases.

\subsection{Information diffusion models}

One way to understand and study the dynamics of OSNs is to build models that represent users interactions. Having a reliable representation enables the conception of simulations that can give support to understand the events that take place in the network.\\

\noindent
\textit{Models paradigms}

In \citet{borge-holthoefer_cascading_2013}, the models used to describe cascades in complex networks are revised. According to them, the models can be divided in two main groups: (a) threshold models and (b) epidemic and rumor models. In both methods, the decision of a user to adopt a certain behavior depends on the neighbors that have already adopted it. In threshold models a user will act only if the proportion of his/her neighbors that are active is superior than a given threshold; in epidemic and rumor models, on the other hand, active users have a probability of infecting each of their neighbors.

An example of the threshold model is provided by \citet{Shakarian2013}, using the model to create a heuristic to identify users able to start a cascade. The method is able to quickly identify a relatively small set of users able to start cascades that cover the whole network, even for large networks with millions of nodes and edges.

Using the epidemic model, we have the work of \citet{gruhl_information_2004} who created a model for information diffusion in blogs, using real data to validate it. They showed that the model faithfully reproduces real behavior, where influential and popular blogs in reality also have relevance in the model's diffusion. \citet{Golub2010} also showed that the epidemic model is an appropriate form of representing cascades, when modeling (the rare\footnote{As noted before, most cascades observed empirically present small depth. However, in \citet{liben-nowell_tracing_2008}, ``large and narrow'' diffusion trees were observed (probably due to the nature of the content being observed -- email chains -- and to the set examined -- successfully diffused chain letters) and were taken as the base structure used on the work of \citet{Golub2010}.}) high depth cascades.

It is important to notice that the epidemic model, based on disease propagation, has its limitations when describing information contagion, given their different nature. One important distinction is the concept of \emph{complex contagion} \citep{centola_complex_2007} which states that, for a behavior be acquired by an individual on social networks, he/she has to be exposed to multiple other individuals. This differs from disease infections, where a single contact with a virus is enough to infect a person (\emph{simple contagion}). \citet{romero_differences_2011} explored this phenomenon on Twitter, showing that multiple exposure to subjects were determinant for contagion. \citet{weng_virality_2013}, however, made a counterpoint showing that although most content spread like complex contagion, some can be properly modeled as simple contagion.

In a different approach, \citet{fapesp2013} built a model where, after collecting behavior data from Twitter, each user receives a probability of posting and a probability for emotions to be expressed. With this, they created a multi-agent model to simulate the behavior of social networks. By building a model based on messages exchanged during United States 2012 presidential campaign, the researchers were able to detect which users were more influential to spread messages. An unexpected conclusion was the fact that the removal of the ten biggest enthusiasts of Barack Obama's campaign would have a larger impact in the network than if Obama himself was removed.\\

\noindent
\textit{Model enhancements}

Some enhancements can be proposed to turn the models more realistic to the OSN context. This is the case of \citet{weng_competition_2012} and \citet{goncalves_modeling_2011}, which considered limitations on the amount of information each user can access and process. This is able to reproduce the fact that many of the information diffusion on OSNs simply lose strength and disappear, regardless the content.

\citet{Gomez2013} discussed ways of modeling and processing information diffusion through multiplex networks. A multiplex network is a network with multiple levels, each level representing a different type of relationship between the network nodes. Therefore, a multiplex network is an adequate model to represent online social networks, as OSN users can be connected in multiple ways (e.g.: different topics may generate different dynamics on the network, creating different diffusion networks connecting users). The proposed analysis revealed relevant aspects of the relationship among those multiple processes.\\

\noindent
\textit{Inferred paths of propagation}

Another area of interest is to determine which are the paths traveled by messages subject to diffusion. \citet{GomezRodriguez2010} were able to infer the order in which users were ``infected'' by a content, by observing the final infected network. By analyzing the timestamps when network nodes shared a content, they calculated the most likely structure that connects the nodes. The algorithm is applied to a large database of blogs' diffusions, achieving high quality results.

\citet{Yang2010} created a method to model and forecast information diffusion, independently of the network structure. For each user of the network, an influence index is estimated, as a measure of the number of users infected by him/her, over time. Thus, for an initial group of infected users, it is possible to predict how many new users will be infected in the future, even without information regarding their connections. Also, the individual influences can be grouped and be used to model the influence dynamics of different classes of users.

\subsection{Influence}
\label{influence}

As already antecipated, another important factor that determines information diffusion in an OSN is the users' capability of influence. An influential user can be determinant to start (or trigger) cascade events, or even change people's opinion and behavior.\\

\noindent
\textit{Locating influential users}

Locating an influential individual in a network is not a trivial task. \citet{cha2010} discussed three metrics aiming to quantify users' influence in OSNs: number of connections (nodes degree), number of mentions, and number of messages reshared (\emph{retweets}) by other users. A discussion of the most appropriate ways to measure influence is done, revealing that simple metrics like number of connections can be misleading to represent the future influence of a user. \citet{weng_twitterrank:_2010} were more optimistic, showing that an adaptation of the PageRank algorithm \citep{page_pagerank_1999} can be used to successfully measure influence on networks.

However, \citet{Bakshy2011}, when analyzing a huge dataset, showed that the theoretical results and metrics are not always confirmed in reality. They discussed that, even though it is possible to identify influential users able to repeatedly start widely scattered cascades, determining a priori which users will influence a cascade process is a hard task. \citet{Borge-Holthoefer2012} also analyzed real data in order to identify influential users from the network topology. Although some influential users are correctly identified in some cases, there are situations where ``badly located'' users are also able to be influential, exceeding expectations.\\

\noindent
\textit{Influence effects}

Researchers have also been interested in evaluating the effects of social influence. \citet{Bakshy2009}, by examining the adoption rate of user-to-user content transfer in Second Life\footnote{On Second Life's virtual world, users are able to share \emph{assets} with other users. An asset can be an ability (e.g.: a dance movement), an item or other customizations.} among friends and strangers, showed that content sharing among known users usually happens sooner than among strangers, although transactions with strangers can influence and reach a wider audience.

\citet{stieglitz_political_2012} analyzed tweets with political opinions and concluded that texts with increased emotional words have stronger influence in the network, being more likely to be shared. \citet{Salathe2012} discussed how the network connections influence opinions and individual sentiment, by observing reactions to a new vaccine campaign in United States. They showed that negative users are more accepted by the network and that users connected with opinionated neighbors tend to be discouraged from expressing opinions.

\subsection{Network's influence on behavior}

Even though individual users have autonomy, it can not be denied that social connections have influence on the formation and evolution of their behaviors and opinions. The OSN analysis enables the empirical observation of the consequences of social connections on individual behavior, and the development of new models and theories capable of explaining those hypothetical associations.\\

\noindent
\textit{Homophily}

A relationship between the topological structure of an OSN and the behavior of its users can be often noticed. In most cases it is not possible to determine what is cause and what is consequence (i.e., if the topology is a result of users behavior, or if the behavior is a consequence of the topology), but the study of one can help in the understanding of the other.

Researchers identified, in general social networks, a tendency that users with common interests are usually connected to each other \citep{mcpherson_birds_2001}. Such phenomenon is called \emph{homophily} and is also verified on OSNs. For example, \citet{bollen_happiness_2011} verified, by investigating the relationship between emotions and social connections, that users considered happy tend to be linked to each other.

\citet{Romero2013} investigated the relationship between the (explicit) network of friendship and the (implicit) network of topical affiliations (i.e., the communities formed by users interested in a common topic). They showed that both networks have considerable intersection (users tend to connect to other users with common interests), such that it is possible to predict friendship from \emph{hashtag} diffusions and also the future popularity of a \emph{hashtag} from the friends network.\\

\noindent
\textit{Users' information processing capability}

\citet{goncalves_modeling_2011} verified whether users are able to surpass, in OSNs, the \emph{Dunbar's number}\footnote{The Dunbar's number is a limit, proposed by the anthropologist Robin Dunbar, for the maximum amount of stable social relationships one person is able to maintain. The actual number usually varies between 100 and 200 and was proposed based on observations of the relation between social group size and brain size in primates \citep{Dunbar1992}.}, given that users usually have hundreds, or even thousands, of connections in such services. After analyzing message exchanges, they showed that, despite the abundance of social connections in OSNs, users are unable to interact regularly with more peers than what is predicted by Dunbar's threshold. \citet{grabowicz_social_2012} studied how the topology affects the type of content transmitted on the network, discussing how users not very close related (intermediary ties) can filter relevant information from several groups, while close relationships (strong ties) can be distracted with a great amount of irrelevant messages.\\

\noindent
\textit{Divergence of opinions in networks}

By examining the information diffusion dynamics on OSN, \citet{romero_differences_2011} studied how users would not immediately adopt an opinion or behavior (such as a new political position) from the first contact with the idea, provided by few initial users. However, if the user is continuously exposed to such content, with many users reinforcing it, the chance of adoption increases. This result is validated on Twitter, where the authors examined how \emph{hashtags} are diffused and the decisive role of multiple exposures.

Based on the relationships established on Twitter, \citet{golbeck_method_2014} estimated the political preferences of users and analyzed how different political opinions coexist in a social network. Also, using the user database together with the predicted political preferences, they were able to analyze the audience of traditional media sources, classifying them as liberal or conservative. This media classification showed to be coherent with previous classification in the literature.

\subsection{Self-organization}

Some research groups studied how users in OSNs, given the absence of central command and their decentralized communication, are able to self-organize in specific situations.\\

\noindent
\textit{Crisis events}

Leysa Palen, Kate Starbird and colleagues \citep{vieweg_microblogging_2010, starbird_chatter_2010, starbird_pass_2010, starbird_voluntweeters:_2011,  Starbird2012} made a deep research on how OSNs can help managing information during crisis events, such as popular uprisings, political protests, natural disasters and humanitarian aid missions. The researchers identified that, among thousands of messages and publications during a crisis, there is the emergence of mechanisms able to deal efficiently with this overload of information. Some of the observed dynamics include the ability of content selection, relevance detection and attribution of roles to specific users. They showed that the largest information cascades during those events tend to happen with important content, being a way to emphasize content worth to be viewed by other users. Also, the network is able to identify reliable users (like on-site witnesses) and give relevance to their posts, by sharing them more often. Thereby, just by observing the content circulating on SNSs, it is possible to quickly identify the most important or urgent information and even coordinate actions in order to help and assist people.
\nocite{vieweg_microblogging_2010, starbird_chatter_2010, Sarcevic2012a, qu_microblogging_2011}\\

\noindent
\textit{Social curating}

Another self-organizing ability of OSNs is content curating, which is the ability of collectively selecting and filtering content relevant to users. This process can happen both spontaneously in traditional SNSs or in dedicated services like \emph{Pinterest}\footnote{\url{http://www.pinterest.com}} or \emph{Tumblr}\footnote{\url{https://www.tumblr.com}}, where users can collaboratively build collections of diverse subjects, selecting content from the Internet.

\citet{Liu2010curating} explored the skills involved in the curating process, describing seven distinct abilities of a social network, namely: collecting, organizing, preserving, filtering, crafting a story, displaying and facilitating discussions. Those skills are compared to actual professional skills (archivist, librarian, preservationist, editor, storyteller, exhibitor, docent, respectively), emphasizing how impressive is the network ability to promote self-organization, being able to specialize and accomplish complex tasks.

\citet{Zhong2013}, in a comprehensive study, described with details the process and the mechanisms of curating in Last.fm\footnote{\url{http://www.lastfm.com}} and Pinterest services, discussing users motivations behind it. They also showed that the social curating process is able to give value to items differently from centralized strategies, being an important source of opinion and measurement of quality. However, the community choices can still be biased, specially when dealing with items already popular in the network, or previously promoted by the service.

\section{Final remarks}
\label{conclusion}

In this work, we performed a comprehensive analysis of research published on online social network analysis, from a Computer Science perspective. Different topics of inquiry were distinguished and a taxonomy was proposed to organize them. For each area, we defined the scope of the works included in it, some of the most representative works, highlighting the discoveries, discussions and challenges of each field.

As seen in the previous sections, computational research in OSN analysis is wide and diverse, enabling the application of techniques from many fields like graph theory, complex networks, dynamic systems, computational simulation, machine learning, natural language processing, data mining, spatio-temporal modeling, among others.

Although many aspects of the presented areas are still being developed, some general movements on the research's course could be identified. The simple characterization of OSN structures, much valued on the first studies, was progressively replaced by studies of users' behavior on the network and the complex dynamic produced by them. Works using social data for different purposes are also very common, with the knowledge extracted being often considered as a valuable representative of human behavior or opinion.\\

\noindent
\textit{Future perspectives}

Predicting the next steps of research on OSN is a challenging and risky task. It is even temerarious to predict if the interest on this topic will still be increasing in years to come. Nonetheless, we will list in the following paragraphs some possibilities of new studies that we believe are worth being explored.

Despite the existance of few works combining information from many social networks, we can notice an increase in the number of theoretical and experimental studies dealing with heterogeneous relationships (e.g.: following, friendship, transportation sharing) from one or more concurrent sources \citep{Gomez-Gardenes2012, Gomez2013, Mucha2010, Sun2012}. This kind of analysis opens several new roads for research, making possible to have a more complete overview of how individuals interact and influence each other, to better track the evolution of a piece of information and to evaluate how specialized may be the use of different social networks -- what may help us to estimate how representative is user behavior in OSNs -- , just to cite a few examples. An important aspect to single out is that this data may also be obtained from sources other than OSNs, like surveys and interviews or sensors (e.g.: GPS in smartphones). In fact, we believe that, with the emergence of the ``Internet of Things'', this offline data will acquire prominence in computational studies about human behavior.

In addition to the use of different sources for context awareness, the deeper understanding of how networks evolve during time is also a likely subject to appear in the future. Most studies still consider the network structure as a fixed object, ignoring its transformation and plasticity. The limitation of current methods may be seen in information diffusion analysis, for instance, as the disregard of when a connection is active may create paths that are not temporally consistent and reduce artificially the distance between individuals. More work is required to understand what are the transformations that take place on each kind of network, their impact on the processes observed in complex systems and how such processes influence the evolution of networks themselves.

The knowledge drawn from online social networks may impact not only computer science, but it may provoke a revolution in social sciences. The burgeoning cross-disciplinary field of computational social science benefits from computational methods, as multi-agent based models, network analysis and machine learning, in order to build a fast, data-driven science. The program of this new data intensive discipline intends to make use of partially structured data available in the Internet, in order to validate and complement existing social theories, or even to propose new research explanations to social phenomena. The use of data from OSNs can not only make much faster the currently time-consuming process of gathering social data, but it may also improve the reproducibility of research in social sciences, as every step of the research -- from data collection to its analysis -- may be audited and reproduced by external agents.\\

\noindent
\textit{Challenges}

Even though the volume of work analyzing OSNs is significant, the area still presents some open challenges, that deserve to be further addressed by researchers.

One initial challenge is associated with the tools and methodologies used. We see that most approaches of OSN studies (specially social data analysis) focus on characteristics of users or messages, but few have a more systemic view, approaching network effects. Therefore, we believe that there is a promising niche to be further explored using methods from complex systems and network science, trying to understand, for instance, the roles of topology, homophily, heterogeneity in individual behaviors and collective cognition in such social systems. This kind of research, however, demands tools and strategies yet to be discovered and experienced. More effort, thus, is required to build a robust theoretical framework to tackle those problems adequately.

After approximately ten years in the spotlight, OSNs are still a topic of interest of general media and academia. Buzzwords like ``social'', ``big data'' and ``complexity'' are increasingly popular and the amount of new scientific papers related to them grows each year. At the same time that more discoveries are made it gets more difficult to properly select relevant works and validate new results presented in literature. One of the main aims of this work is, precisely, to help researchers with the task of organizing and selecting material on OSN analysis.

The lack of ethical considerations in most of the observed works is left as our final remark. Even though we focused on computational approaches to online social networks, the information collected and the knowledge produced by the works we analyzed have direct implications on societies. For example, the theories and methods developed in this research area can, potentially, be used in harmful ways by authoritarian regimes or abusive advertising campaigns. Privacy is also an important issue as, by analyzing public data and behaviors in OSNs, data scientists may uncover implicit information about specific individuals, information that such individuals may have never intended to made public. As OSN analysis is a strongly interdisciplinary field, we believe that this is a current challenge, indispensable to be considered.

\section*{Acknowledgements}

The authors sincerely thank Romis Attux, Leonardo Maia and Fabrício Olivetti de França for their kind effort of revising this work and contributing with corrections and new insights.

Part of the results presented in this work were obtained through the project ``Training in Information Technology'', funded by Samsung Eletronics of Amazonia LTDA., using resources from Law of Informatics (Brazilian Federal Law Number 8.248/91).

\bibliographystyle{plainnat}
\bibliography{survey.bib}

\end{document}